\documentstyle[12pt]{ioplppt}
\newtheorem{theorem}{Theorem}[section]
\newtheorem{proposition}[theorem]{Proposition}
\newtheorem{lemma}[theorem]{Lemma}
\eqnobysec
\begin{document}
\jl{1}
\title{Solutions of a discretized Toda field equation 
for $D_{r}$ from Analytic Bethe Ansatz}[Solutions
 of a discritized Toda field equation]
 
\author{Zengo Tsuboi and Atsuo Kuniba\ftnote{1}{E-mail: 
atsuo@hep1.c.u-tokyo.ac.jp}
}

\address{Institute of Physics, University of Tokyo \\
 Komaba 3-8-1, Meguro-ku, Tokyo 153 Japan}

\begin{abstract}
Commuting transfer matrices of $U_{q}(X_{r}^{(1)})$ vertex models 
obey the functional relations which can be viewed as an $X_{r}$ 
type Toda field equation on discrete space time. Based on analytic 
Bethe ansatz we present, for $X_{r}=D_{r}$, a new expression of 
its solution in terms of determinants and Pfaffians.
\end{abstract}
\maketitle

\section{Introduction}
In [KNS], a family of functional relations, $T$-system, 
was proposed for commuting transfer matrices of solvable 
lattice models associated to any quantum affine algebras
$U_q(X^{(1)}_r)$.
For $X_r = D_r$ it reads as follows:
\numparts 
\begin{eqnarray}
\fl      T_{m}^{(a)}(u-1) T_{m}^{(a)}(u+1)  = 
    T_{m+1}^{(a)}(u) T_{m-1}^{(a)}(u)+
        T_{m}^{(a-1)}(u) T_{m}^{(a+1)}(u) \quad 1 \le a \le r-3,
        \label{t-sys1} \\
\fl     T_{m}^{(r-2)}(u-1) T_{m}^{(r-2)}(u+1)  =  T_{m+1}^{(r-2)}(u) 
        T_{m-1}^{(r-2)}(u)+
        T_{m}^{(r-3)}(u) T_{m}^{(r-1)}(u) T_{m}^{(r)}(u),
        \label{t-sys2} \\
\fl     T_{m}^{(a)}(u-1) T_{m}^{(a)}(u+1)  =  
    T_{m+1}^{(a)}(u) T_{m-1}^{(a)}(u)+
        T_{m}^{(r-2)}(u) \qquad a=r-1, r.
        \label{t-sys3}
\end{eqnarray} 
\endnumparts
where $T^{(a)}_m(u)$ $(m \in {\bf Z}, 
u \in {\bf C}: \hbox{ spectral parameter})$
denote the transfer matrices with the auxiliary space
labeled by $a$ and $m$.
We shall employ the boundary condition 
$T_{-1}^{(a)}(u) = 0, \, T^{(a)}_0(u) = 1$, which
is natural for the transfer matrices.
Then, solving (1.1) successively, 
one can express $T^{(a)}_m(u)$ uniquely 
as a polynomial of the fundamental ones
$T^{(1)}_1, \ldots, T^{(r)}_1$.
The aim of this paper is to give a new expression to the solution 
of (1.1) motivated by the analytic Bethe ansatz [R].
There is an earlier solution in [KNH], which are
expressed only by the fundamental ones
$T^{(1)}_1, \ldots, T^{(r)}_1$.
However, in this paper we begin by introducing 
the auxiliary 
transfer matrix (or `dress function' in the analytic Bethe ansatz)
${\cal T}^a(u)$ (\ref{calT}) for any $a \in {\bf Z}$
and establish a new functional relation as in proposition\ref{propo}.
For $1 \le a \le r-2$
${\cal T}^a(u)$ is just $T^{(a)}_1(u)$ while
for $a \ge r-1$ it is quadratic in $T^{(r)}_1$ 
and $T^{(r-1)}_1$. 
We then express the solution as the
determinants and Pfaffians with 
matrix elements $0, \pm {\cal T}^a, 
\pm T^{(r-1)}_1$ or $\pm T^{(r)}_1$.
Moreover those determinants and Pfaffians are taken over
the matrices with dense distributions of non zero elements
as opposed to the sparse ones in [KNH].

The two types of the representations of the 
solutions obtained here and in [KNH] are 
significant in their own right.
The sparse type one [KNH]
arises straightforwardly from a manipulation of the 
$T$-system only.
On the other hand, the dense type one
is more connected with the analytic Bethe ansatz idea [KS],
in view of which it is most natural to introduce the ${\cal T}^a$ as
well as the $Q$-functions 
$Q_1(u), \ldots, Q_r(u)$.
It should be noted that $T^{(a)}_m(u)$ 
in this paper is a solution of (1.1) 
for arbitrary $Q$-functions.
The definition through the 
Bethe equations as in (2.1-2) is needed only when 
one requires $T^{(a)}_m(u)$ to
yield the actual transfer matrix spectra.
We note that similar two such representations are
available also for the solution of $B_r$ $T$-system 
in [KOS] and [KNH].

As the previous cases [KOS, KNH],
all the proofs of the determinant and Pfaffian formulae
reduce essentially
to the Jacobi identity (\ref{jacobi}), a well known 
machinery in soliton theories.
In fact, it was 
firstly pointed out in [KOS] that the $T$-system for 
$U_q(X^{(1)}_r)$ 
may be viewed as a Toda field equation 
[LS,MOP] with discrete space time variables $u$ and $m$.
Mathematically, it implies a common structure 
between discretized soliton equations 
(cf [AL,DJM,H]) and 
representation rings of finite dimensional
modules over Yangians or quantum affine algebras.
Our new solution here
exemplifies such an interplay further.
See [KLWZ] for a similar perspective.

The outline of the paper is as follows.
Section 2 is a brief review of the $D_r$ case of the 
analytic Bethe ansatz results [KS].
For $T^{(1)}_1(u), T^{(r)}_1(u)$ and $T^{(r-1)}_1(u)$
it partially overlaps the earlier result in [R].
Proposition\ref{propo} is new and plays a key role in the 
subsequent arguments.
In section 3 we present the solution, which is proved 
in section 4.
Section 5 is devoted to discussion.
Appendix A provides a number of formulae similar to those
used in section 4.
\section{Review of the results for fundamental representations}
In this section we basically follow [KS].
Let $\{\alpha_1,\dots,\alpha_r \}$ be the simple roots normalized so
 that $(\alpha_a | \alpha_b)=$ Cartan matrix.  
 The Bethe ansatz equation reads :
\begin{eqnarray}
        -1=\prod_{b=1}^{r}\frac{Q_{b}(u_k^{(a)}+(\alpha_a|\alpha_b))}
           {Q_{b}(u_k^{(a)}-(\alpha_a|\alpha_b))} \qquad 
           1\le a \le r,\ 1\le k\le N_{a},  \\
        Q_{a}(u)=\prod_{j=1}^{N_{a}}[u-u_j^{(a)}],
        \label{Q_a}
\end{eqnarray}
where $[u]=(q^u-q^{-u})/(q-q^{-1})$ and $N_{a} \in {\bf Z }_{\ge 0}$.
In this paper, we suppose that $q$ is generic. 
We define a set 
\begin{equation}
        J=\{ 1,2,\dots,r,\bar{r},\dots,\bar{1} \} 
\end{equation}
with the partial order 
\begin{eqnarray}
        1\prec 2 \prec \cdots \prec r-1 \prec \left.
                                          \begin{array}{c}
                                          r \\
                                          \overline{r} \\
                                          \end{array}
                                    \right.
    \prec \overline{r-1} \prec \cdots \prec \bar{2} \prec \bar{1}. 
\end{eqnarray}
Note that there is no order between $r$ and $\bar{r}$.
For $a \in J $, set
\begin{eqnarray}
\fl     z(a;u)=\frac{Q_{a-1}(u+a+1)Q_{a}(u+a-2)}{Q_{a-1}(u+a-1)Q_{a}(u+a)} 
    \qquad {\rm for} \quad 1 \le a \le r-2,
        \nonumber \\
\fl     z(r-1;u)=\frac{Q_{r-2}(u+r)Q_{r-1}(u+r-3)Q_{r}(u+r-3)}
        {Q_{r-2}(u+r-2)Q_{r-1}(u+r-1)Q_{r}(u+r-1)},
        \nonumber \\
\fl  z(r;u)=\frac{Q_{r-1}(u+r+1)Q_{r}(u+r-3)}
        {Q_{r-1}(u+r-1)Q_{r}(u+r-1)}, \nonumber \\ 
\fl     z(\bar{r};u)=\frac{Q_{r-1}(u+r-3)Q_{r}(u+r+1)}
        {Q_{r-1}(u+r-1)Q_{r}(u+r-1)},
         \label{def-z} \\
\fl     z(\overline{r-1};u)=\frac{Q_{r-2}(u+r-2)Q_{r-1}(u+r+1)Q_{r}(u+r+1)}    
        {Q_{r-2}(u+r)Q_{r-1}(u+r-1)Q_{r}(u+r-1)}, 
        \nonumber \\
\fl     z(\bar{a};u)=\frac{Q_{a-1}(u+2r-a-3)Q_{a}(u+2r-a)}
        {Q_{a-1}(u+2r-a-1)Q_{a}(u+2r-a-2)} 
    \qquad {\rm for} \quad 1 \le a \le r-2,
         \nonumber
\end{eqnarray}
where $Q_{0}(u)=1$. For $(\xi_{1},\dots,\xi_{r}) \in 
\{ \pm \}^{r}$, we define the function $sp(\xi_{1},\dots,\xi_{r};u)$ 
recursively by 
\begin{eqnarray}
        sp(+,+,\xi_{3},\dots,\xi_{r};u)=
        \tau^{Q}sp(+,\xi_{3},\dots,\xi_{r};u),
        \nonumber \\
    sp(+,-,\xi_{3},\dots,\xi_{r};u)=\frac{Q_{1}(u+r-3)}{Q_{1}(u+r-1)}
        \tau^{Q}sp(-,\xi_{3},\dots,\xi_{r};u),
        \nonumber \\
    sp(-,+,\xi_{3},\dots,\xi_{r};u)=\frac{Q_{1}(u+r+1)}{Q_{1}(u+r-1)}
        \tau^{Q}sp(+,\xi_{3},\dots,\xi_{r};u+2),
        \nonumber \\
        sp(-,-,\xi_{3},\dots,\xi_{r};u)=
        \tau^{Q}sp(-,\xi_{3},\dots,\xi_{r};u+2),
        \label{def-sp}
\end{eqnarray}
with the following initial conditions:
\begin{eqnarray}
   sp(+,+,+,+;u)=\frac{Q_{4}(u-1)}{Q_{4}(u+1)}, \nonumber \\ 
    sp(+,+,-,-;u)=\frac{Q_{2}(u)Q_{4}(u+3)}{Q_{2}(u+2)Q_{4}(u+1)},
          \nonumber \\
        sp(+,-,+,-;u)=\frac{Q_{1}(u+1)Q_{2}(u+4)Q_{3}(u+1)}
                       {Q_{1}(u+3)Q_{2}(u+2)Q_{3}(u+3)},
        \nonumber \\
        sp(+,-,-,+;u)=\frac{Q_{1}(u+1)Q_{3}(u+5)}{Q_{1}(u+3)Q_{3}(u+3)},
        \nonumber \\ 
        sp(-,+,+,-;u)=\frac{Q_{1}(u+5)Q_{3}(u+1)}{Q_{1}(u+3)Q_{3}(u+3)},
         \label{ini-condition} \\
        sp(-,+,-,+;u)=\frac{Q_{1}(u+5)Q_{2}(u+2)Q_{3}(u+5)}
                       {Q_{1}(u+3)Q_{2}(u+4)Q_{3}(u+3)},
           \nonumber  \\
        sp(-,-,+,+;u)=\frac{Q_{2}(u+6)Q_{4}(u+3)}{Q_{2}(u+4)Q_{4}(u+5)},
        \nonumber \\ 
        sp(-,-,-,-;u)=\frac{Q_{4}(u+7)}{Q_{4}(u+5)}.
        \nonumber       
\end{eqnarray}
Here $\tau^{Q}$ is the operation $Q_{a}\mapsto Q_{a+1}$, that is, 
\begin{eqnarray}
  \tau^{Q}f(Q_{1}(u+x_{1}^1),Q_{1}(u+x_{2}^1),\dots,
  Q_{2}(u+x_{1}^2),Q_{2}(u+x_{2}^2),\dots)
        \nonumber \\
  =f(Q_{2}(u+x_{1}^1),Q_{2}(u+x_{2}^1),\dots,
  Q_{3}(u+x_{1}^2),Q_{3}(u+x_{2}^2),\dots)
\end{eqnarray}
for any function $f$.
We shall use the functions ${\cal T}^{a}(u)$  for $a \in {\bf Z }$ 
and $u \in {\bf C }$ determined by the generating series
\begin{eqnarray}
\fl     (1+z(\bar{1};u)X)\cdots (1+z(\overline{r-1};u)X)[-1+(1+z(r;u)X)
        (1-z(\bar{r};u)Xz(r;u)X)^{-1} \nonumber \\
\fl     +(1+z(\bar{r};u)X)(1-z(r;u)X
        z(\bar{r};u)X)^{-1}](1+z(r-1;u)X)\cdots (1+z(1;u)X) \nonumber \\
\fl     =\sum_{a=-\infty}^{\infty} {\cal T}^{a}(u+a-1)X^{a}, 
        \label{generating}
\end{eqnarray} 
where $X$ is a shift operator $X=\e^{2\partial_{u}}$.
Namely, for $a<0$ ${\cal T}^{a}(u)=0$ and for $a \ge0 $, 
\begin{eqnarray}
\fl {\cal T}^{a}(u)=\sum_{{\tiny 
                             \begin{array}{c}
                             i_{1}< \dots < i_{k}, \\ 
                             j_{1}< \dots < j_{l}, \\
                             k+l+2n=a
                             \end{array}
                           }
                     }
z(i_1;v_1) \cdots z(i_k;v_k) \label{calT} \\ \nonumber 
\times z(\overline{r};v_{k+1})z(r;v_{k+2}) \cdots 
z(\overline{r};v_{k+2n-1})z(r;v_{k+2n}) z(\overline{j_l};v_{k+2n+1}) 
\cdots z(\overline{j_1};v_a) ,
\end{eqnarray}
where $i_{\alpha},j_{\beta}\in \{1,\dots , r\}$, 
$k,l,n \in {\bf Z}_{\ge 0}$ and $v_{\gamma}=u+a-2\gamma+1$.
We define the functions $T_{1}^{(a)}(u)$ for $1 \le a \le r$ that 
correspond to the dress parts of eigenvalue of transfer matrix in 
[KS].
\numparts
\begin{eqnarray}
        T_{1}^{(a)}(u)={\cal T}^a(u) \qquad {\rm for} \quad 1 \le a\le r-2,
        \label{defT^a} \\
    T_{1}^{(r-1)}(u)=\sum_{(\xi_1, \ldots, \xi_r) \in Spin^-}
    sp(\xi_{1},\dots,\xi_{r};u),
        \label{def-sp+} \\
        T_{1}^{(r)}(u)=\sum_{(\xi_1, \ldots, \xi_r) \in Spin^+}
    sp(\xi_{1},\dots,\xi_{r};u),
        \label{def-sp-}
\end{eqnarray}
\endnumparts 
where for $\epsilon=\pm$ we have put
\begin{equation}
Spin^\epsilon = \{(\xi_1, \ldots, \xi_r) : \xi_{j}=\pm , 
\prod_{j=1}^{r}\xi_{j} = \epsilon \}. \label{spin}
\end{equation}
\begin{theorem} {\rm ([KS])} 
For any integer $a$, ${\cal T}^a(u)$, $T_{1}^{(r-1)}(u)$ and 
$T_{1}^{(r)}(u)$ are free of poles under the condition that
the BAE is valid.   
\end{theorem}
Actually in [KS] only $a \le r-2$ case was considered for ${\cal T}^a(u)$
but the proof therein is valid for any $a$.
Except this theorem, all the definitions and the statements
in this paper make sense without assuming (2.1-2) as mentioned in the 
introduction.

Let us now explain the relations between $z(a;u)$ and 
$sp(\xi_{1},\dots,\xi_{r};u)$. 
Define $i_1 < \dots < i_k$, $I_1 <\dots <I_{r-k}$ $(0 \le k \le r)$ and 
$j_1 < \dots < j_l$, $J_1 < \dots < J_{r-l}$  $(0 \le l \le r)$ 
using the two sequences $(\xi_{1},\dots,\xi_{r})$ 
and $(\eta_{1},\dots,\eta_{r})  \in \{ \pm \}^{r}$ as follows 
\begin{eqnarray}
        \xi_{i_1}=\cdots =\xi_{i_k}=+, \xi_{I_1}=\cdots =\xi_{I_{r-k}}=-, 
        \nonumber \\
        \eta_{j_1}=\cdots =\eta_{j_l}=-, \eta_{J_1}=\cdots =\eta_{J_{r-l}}=+.
        \label{relation-ineq}
\end{eqnarray}
Using the relations (\ref{def-z}), (\ref{def-sp}), 
(\ref{ini-condition}) and induction on $r$, we have 
\begin{proposition}
For any $ a \in {\bf Z }_{\ge 0} $, 
\numparts
\begin{eqnarray}
        \prod_{n=1}^{a} z(b_{n};u+a+1-2n) 
        \nonumber \\
        =sp(\xi_{1},\dots,\xi_{r};u-r+a+1)
        sp(\eta_{1},\dots,\eta_{r};u+r-a-1),
        \label{z=spsp1} \\
        b_{n}=
        \left\{
          \begin{array}{ll}
            i_{n} &
            {\rm for} \quad 1 \le n \le k \\
            r &
            {\rm for} \quad k < n \le a-l \quad and 
            \quad n \equiv k \bmod{2} \\
            \bar{r} &
            {\rm for} \quad k < n \le a-l \quad and 
            \quad n \not\equiv k \bmod{2}  \\
            \overline{j_{a+1-n}} &
            {\rm for} \quad a-l < n \le a \\
            \end{array}
    \right.
\end{eqnarray}
\endnumparts 
 if $k+l \le a$ and $a \equiv l+k \bmod{2}$. \\ 
For any $a \in {\bf Z }_{\le 2r-2}$,
\numparts
\begin{eqnarray}
        \prod_{n=1}^{2r-a-2} z(b_{n}^{\prime} ;u+2r-a-1-2n)
        \nonumber \\
        =sp(\xi_{1},\dots,\xi_{r};u-r+a+1)
        sp(\eta_{1},\dots,\eta_{r};u+r-a-1) ,
        \label{z=spsp2} \\
\fl     b_{n}^{\prime} =
     \left\{
          \begin{array}{ll} 
            J_{n} & 
            {\rm for} \quad 1 \le n \le r-l \\
            r &
            {\rm for} \quad r-l < n \le r+k-a-2  
            \quad and 
            \quad n-r+l\equiv 0 \bmod{2} \\
            \bar{r} &
             {\rm for} \quad r-l < n \le r+k-a-2  
            \quad and 
            \quad n-r+l\equiv 1 \bmod{2} \\
            \overline{I_{2r-a-1-n}} & 
            {\rm for} \quad r+k-a-2 < n \le 2r-a-2 \\
            \end{array}
    \right.
\end{eqnarray}
\endnumparts
 if $k+l \ge a+2$ and $a \equiv l+k \bmod{2}$.
\end{proposition}
For $a \le r-2$, (\ref{z=spsp1}) is eq. (B.1) in [KS]. 
The following new functional relation is the $D_{r}$ version of 
eq. (2.14) in [KOS], which is derived by summing up the equations 
(\ref{z=spsp1}) and (\ref{z=spsp2}). 
\begin{proposition}
\begin{eqnarray}
\fl {\cal T}^{a}(u)+{\cal T}^{2r-a-2}(u) \label{fun.rel.} 
 &=&T_{1}^{(r)}(u+r-a-1)T_{1}^{(r-\delta_{r-a})}(u-r+a+1) \nonumber  \\
&+&T_{1}^{(r-1)}(u+r-a-1)T_{1}^{(r-\delta_{r-a-1})}(u-r+a+1),\label{key}
\end{eqnarray}
where 
\[ \delta_{i}= 
     \left\{
            \begin{array}{ll}
            0 & if \quad i \in 2{\bf Z } \\
            1 & if \quad i \in 2{\bf Z }+1 . 
            \end{array}
    \right.
 \]
 \label{propo}
\end{proposition}
Note that (\ref{fun.rel.}) is invariant under the exchange
 $a \leftrightarrow  2r-2-a$, in paticular 
\begin{equation}
        {\cal T}^{r-1}(u)=T_{1}^{(r)}(u)T_{1}^{(r-1)}(u).
\end{equation}
\section{Main results}
Set  
\begin{eqnarray}
 x_{j}(u)=T_{1}^{(r-\delta_{j-1})}(u+2j-2), \quad 
y_{j}(u)=T_{1}^{(r-\delta_{j})}(u+2j-2) ,\nonumber \\   
t_{ij}(u)=T_{1}^{(r+i-j-1)}(u+i+j-2), \quad 
a_{ij}(u)=x_{i}(u)y_{j}(u)-t_{ij}(u), \\ 
b_{ij}(u)=y_{i}(u)x_{j}(u)-t_{ij}(u). \nonumber 
\end{eqnarray}
By the definition one has 
\begin{eqnarray}
\fl     x_{i}(u+2)&=&y_{i+1}(u), \quad y_{i}(u+2)=x_{i+1}(u),\quad  
        t_{ij}(u+2)=t_{i+1 \; j+1}(u), \nonumber \\
\fl     a_{ij}(u+2)&=&b_{i+1 \; j+1}(u), \qquad 
        b_{ij}(u+2)=a_{i+1 \; j+1}(u) . 
        \label{relation-sift}
\end{eqnarray}
Now we introduce $(m+1)\times (m+1)$ matrix 
${\cal S}_{m+1}(u)=({\cal S}_{ij})_{1\le i,j\le m+1}$ whose $(i,j)$ 
elements are given by 
\begin{eqnarray}
 \fl {\cal S}_{ij} =
    \left\{
          \begin{array}{ll}
            0 & 
            {\rm for} \quad i=j=1 \\
            T_{1}^{(r-\delta_{j})}(u+2j-4) & 
            {\rm for} \quad  i=1 \quad and \quad 2 \le j \le m+1 \\
            -T_{1}^{(r-\delta_{i})}(u+2i-4) & 
            {\rm for} \quad 2 \le i \le m+1 \quad and \quad j=1 \\
            -{\cal T}^{r+i-j-1}(u+i+j-4) & 
            {\rm for} \quad 2 \le i, j \le m+1.
      \end{array}
    \right.
\end{eqnarray}
Using the relation (\ref{defT^a}) and (\ref{fun.rel.}), the matrix 
elements of $ {\cal S}_{m+1}(u) $ can be rewritten as
\begin{eqnarray}
 \fl {\cal S}_{ij} =
    \left\{
          \begin{array}{ll}
            0 & 
            {\rm for} \quad i=j=1 \\
            x_{j-1} & 
            {\rm for} \quad  i=1 \quad and \quad 2 \le j \le m+1 \\
            -x_{i-1} & 
            {\rm for} \quad 2 \le i \le m+1 \quad and \quad j=1 \\
            -x_{i-1} y_{i-1} & 
            {\rm for} \quad i=j \quad and \quad 2 \le i \le m+1 \\
            -t_{i-1 \; j-1} & 
            {\rm for} \quad 2\le i < j \le m+1 \\
            t_{j-1 \; i-1}-x_{j-1}y_{i-1}-x_{i-1}y_{j-1} &  
            {\rm for} \quad 2 \le j < i \le m+1.
      \end{array}
    \right.
\end{eqnarray}
For $ {\cal S}_{m+1}(u)=( {\cal S}_{ij})_{1 \le i, j \le m+1}$, 
we introduce the following anti-symmetric matrices
 $ {\cal C}_{m+1}(u) $ and $ {\cal R}_{m+1}(u)$ :
\numparts 
\begin{eqnarray}
  {\cal C}_{m+1}(u) :={\cal S}_{m+1}(u) \prod_{j=2}^{m+1}
   P(1,j;-y_{j-1}(u)) 
    \nonumber \\
 =\left(
   \begin{array}{ccccc}
        0      & x_{1}   & x_{2}  & \ldots & x_{m}  \\
        -x_{1} & 0       & a_{12} & \ldots & a_{1m}  \\
        -x_{2} & -a_{12} & 0      & \ldots & a_{2m}  \\       
        \vdots & \vdots  & \vdots & \ddots & \vdots  \\
        -x_{m} & -a_{1m} & -a_{2m}& \ldots & 0
   \end{array} 
   \right), \\ 
{\cal R}_{m+1}(u) :=\left( \prod_{i=2}^{m+1} P(i,1;y_{i-1}(u)) \right) 
{\cal S}_{m+1}(u)
 \nonumber \\ 
 =\left(
   \begin{array}{ccccc}
        0      & x_{1}   & x_{2}  & \ldots & x_{m}  \\
        -x_{1} & 0       & b_{12} & \ldots & b_{1m}  \\
        -x_{2} & -b_{12} & 0      & \ldots & b_{2m}  \\
        \vdots & \vdots  & \vdots & \ddots & \vdots  \\
        -x_{m} & -b_{1m} & -b_{2m}& \ldots & 0
   \end{array}
   \right).
\end{eqnarray}
\endnumparts 
Here 
\begin{equation}
P(i,j;c)=E+cE_{ij}
\end{equation}
is the $m+1$ by $m+1$ matrix with $E$ the identity and 
$E_{i j}$ the matrix unit.
The products of $P$'s in the above are commutative.
For any matrix ${\cal M}(u)$, we shall let 
${\cal M}\left[
    \begin{array}{ccc}
          i_{1} & \dots & i_{k}  \\
          j_{1} & \dots & j_{k}
    \end{array}
  \right](u)$
denote the minor matrix removing 
$i_{l} $\symbol{39}s rows and $j_{l} 
$\symbol{39}s columns from ${\cal M}(u)$.
Our main results in this paper is given as follows: 
\begin{theorem} \label{main-th}
The following determinant and Pfaffian expressions solve the  
$D_{r}$ $T$-system (1.1).
\numparts 
\begin{eqnarray}
\fl    T_{m}^{(a)}(u)=
       \det_{1\le i,j \le m}[{\cal T}^{a+i-j}(u+i+j-m-1)] 
       \quad {\rm for} 
       \ a \in \{ 1,2,\dots,r-2 \}, 
       \ m \in {\bf Z }_{\ge 0},
      \label{T_m^(a)} \\
\fl     T_{m}^{(r)}(u)=
           \left\{
           \begin{array}{ll}
            {\rm pf}[{\cal C}_{m+1}
              \left[
              \begin{array}{c}
               1 \\
               1
              \end{array}
              \right]
              (u-m+1)] & 
              {\rm for} \quad  m \in 2{\bf Z }_{\ge 0} \\
            {\rm pf}[{\cal C}_{m+1}(u-m+1)] & 
             {\rm for} \quad  m \in 2{\bf Z }_{\ge 0}+1,
            \end{array}
           \right.
        \label{T_m^(r)} \\
\fl     T_{m}^{(r-1)}(u)=
           \left\{
           \begin{array}{ll}
            {\rm pf}[{\cal C}_{m+2}
              \left[
              \begin{array}{cc}
               1 & 2\\
               1 & 2
              \end{array}
              \right]
              (u-m-1)] & 
             {\rm for} \quad  m \in 2{\bf Z }_{\ge 0} \\
            {\rm pf}[{\cal C}_{m+2}
            \left[
              \begin{array}{c}
               2 \\
               2
              \end{array}
              \right]
            (u-m-1)] &
            {\rm for} \qquad  m \in 2{\bf Z }_{\ge 0}+1.
            \end{array}
           \right.
        \label{T_m^(r-1)} 
\end{eqnarray}
\endnumparts
\end{theorem}

\section{Proof of Theorem3.1 }
At first, we present a number of lemmas that are necessary for the
 proof.
The following Jacobi identity $(b \neq c)$ plays an important role in 
this section.
\begin{equation}
\fl \det {\cal M}\left[
   \begin{array}{c}
        b \\
        b 
   \end{array}
  \right]
 \det {\cal M}
   \left[
   \begin{array}{c}
        c \\ 
        c 
   \end{array}
  \right]-
 \det {\cal M}\left[
   \begin{array}{c}
        b \\
        c 
   \end{array}
  \right]
 \det {\cal M}\left[
   \begin{array}{c}
        c \\
        b 
   \end{array}
  \right]=
 \det {\cal M}\left[
   \begin{array}{cc}
        b & c\\
        b & c
   \end{array}
  \right]
   \det {\cal M}.     
        \label{jacobi}
\end{equation}
\begin{lemma} \label{formallemma} ([KNS])
For any $a, m \in {\bf Z }_{\ge 0}$ and $ u \in {\bf C } $ put
\begin{equation}
\fl    {\cal T}_{m}^{a}(u)=
       \det_{1\le i,j \le m}[{\cal T}^{a+i-j}(u+i+j-m-1)] .
\end{equation}
Then the following functional relation is valid.
\begin{equation}
\fl     {\cal T}_{m}^{a}(u-1) {\cal T}_{m}^{a}(u+1)  = 
    {\cal T}_{m+1}^{a}(u) {\cal T}_{m-1}^{a}(u)+
        {\cal T}_{m}^{a-1}(u) {\cal T}_{m}^{a+1}(u).
        \label{formal relation} 
\end{equation}
\end{lemma}
Proof. \\ 
Apply (\ref{jacobi}) for $(b,c)=(1,m+1)$ to
${\cal M}= [{\cal T}^{a+i-j}(u+i+j-m-2)]_{1\le i,j \le m+1} $.
 \fullsqr   
\begin{lemma} \label{lemma-t-sys2.2}
For (\ref{T_m^(a)})-(\ref{T_m^(r-1)}) to satisfy
(\ref{t-sys2}) it is enough to show 
\begin{equation}
\fl  T_{m}^{(r-1)}(u)T_{m}^{(r)}(u)={\cal T}_{m}^{r-1}(u)
\label{t-sys2.2}
\end{equation}
\end{lemma}
Proof. From Lemma \ref{formallemma} and (\ref{T_m^(a)}), we have 
$T_{m}^{(a)}(u) = {\cal T}_{m}^a(u)$ for $1\le a \le r-2 $.
Then compare (\ref{t-sys2}) and (\ref{formal relation}) for $a=r-2$.
\fullsqr \\
By noting  $\det[P(i,j;c)]=1$, we have 
\begin{lemma}
\quad \\ 
\begin{equation}
\fl  \det[{\cal S}_{m+1}(u)]
  =\det[{\cal C}_{m+1}(u)]
  =\det[{\cal R}_{m+1}(u)]      .
        \label{relation-CR}
\end{equation} 
\end{lemma}  
We shall further need
\begin{lemma} \label{main-lemma}
For $m \in {\bf Z }_{\ge 0}$, $T_m^{(r-1)}(u)$ (\ref{T_m^(r-1)}) and 
 $T_m^{(r)}(u)$ (\ref{T_m^(r)}) satisfy the following relations:
\numparts 
\begin{eqnarray}
 \fl T_{m}^{(r-1)}(u+1) T_{m-1}^{(r)}(u)=
    \left\{
          \begin{array}{ll}
            \det[{\cal S}_{m+1}
            \left[
              \begin{array}{c}
                m+1 \\
                1
              \end{array}
            \right]
            (u-m+2)] & 
            {\rm for} \quad  m \in 2{\bf Z }_{\ge 0} \\
             \det[{\cal S}_{m+2}
            \left[
              \begin{array}{cc}
               2 & m+2 \\
               1 & 2
              \end{array}
            \right]
            (u-m)] & {\rm for} \quad  m \in 2{\bf Z }_{\ge 0} +1,
      \end{array}
    \right.
 \label{relation3}
\end{eqnarray}
\begin{eqnarray}
\fl  T_{m}^{(r-1)}(u) T_{m}^{(r)}(u)=
    (-1)^m \det[{\cal S}_{m+1} 
           \left[
            \begin{array}{c}
             1 \\
             1
            \end{array}
           \right]
           (u-m+1)],
 \label{relation1}
\end{eqnarray}
\begin{eqnarray}
\fl  T_{m}^{(r-1)}(u+1) T_{m+1}^{(r)}(u)=
    (-1)^{m+1} \det[{\cal S}_{m+2}
           \left[
            \begin{array}{c}
             1 \\
             2
            \end{array}
           \right]
           (u-m)],
 \label{relation4}
\end{eqnarray}
\begin{eqnarray}
\fl  T_{m-1}^{(r-1)}(u) T_{m}^{(r-1)}(u+1)=
 (-1)^{m} \det[{\cal S}_{m+2}
           \left[
            \begin{array}{cc}
             1 & 2 \\
             2 & m+2
            \end{array}
           \right]
           (u-m)],
 \label{relation6}
\end{eqnarray}
\begin{eqnarray}
\fl  T_{m-1}^{(r)}(u-1) T_{m}^{(r)}(u)=
    (-1)^{m} \det[{\cal S}_{m+1}
           \left[
            \begin{array}{c}
             1 \\
             m+1
            \end{array}
           \right]
           (u-m+1)],
 \label{relation7}
\end{eqnarray}
\begin{eqnarray}
\fl  T_{m}^{(r-1)}(u) T_{m-1}^{(r)}(u+1)= 
    (-1)^{m} \det[{\cal S}_{m+2}
           \left[
            \begin{array}{cc}
             1 & 2 \\
             2 & 3
         \end{array}
           \right]
           (u-m-1)].
 \label{relation8}
\end{eqnarray}
\endnumparts
\end{lemma}
Proof. \\
All the relations in Lemma \ref{main-lemma} reduces to the 
Jacobi identity.  
First we prove (\ref{relation3})  for $m \in 2{\bf Z }_{\ge 0} $. 
Let ${\cal M}={\cal R}_{m+1}(u-m+2)$ and noting the 
relation (\ref{relation-CR}), we have 
\begin{eqnarray}
\fl     \det {\cal M}=\det[{\cal R}_{m+1}(u-m+2)]=0, \quad 
        \det {\cal M}\left[
            \begin{array}{c}
             1\\
             1 
            \end{array}
      \right] 
    =(T_{m}^{(r-1)}(u+1))^2,  
         \nonumber  \\
        \det {\cal M}\left[
            \begin{array}{c}
             m+1\\
             m+1 
            \end{array}
      \right] 
     = (T_{m-1}^{(r)}(u))^2,   \\
\fl      \det {\cal M}\left[
            \begin{array}{c}
             1 \\
             m+1 
            \end{array}
      \right] =
         \det {\cal M}\left[
            \begin{array}{c}
             m+1 \\
             1 
            \end{array}
      \right]  
     =\det[{\cal S}_{m+1}
          \left[
            \begin{array}{c}
             m+1 \\
             1
            \end{array}
      \right]
        (u-m+2)].  \nonumber
\end{eqnarray}
The first identity follows from the fact that the determinant of 
antisymmetric matrix of odd size should vanish. 
And other ones follow from (\ref{T_m^(r)}), (\ref{T_m^(r-1)}) and 
(\ref{relation-sift}). Substituting  these identity into 
(\ref{jacobi}) for $ (b,c)=(1,m+1)$, we have 
\begin{equation}
  (T_{m}^{(r-1)}(u+1)T_{m-1}^{(r)}(u))^2 
  =(\det[{\cal S}_{m+1}
          \left[
            \begin{array}{c}
             m+1 \\
             1
            \end{array}
      \right]
        (u-m+2)])^2.
        \label{rel-3*3}
\end{equation}
Taking square root of (\ref{rel-3*3}), we have  (\ref{relation3}) for 
$m\in 2{\bf Z }_{\ge 0}$. Relative sign can be determined so that  
the equation is valid for $m=0$ and $2$ or more rigorously, by 
comparing the sign of the coefficient of $x_{1}(u-m+2)\cdots 
 \cdot x_{m}(u-m+2) \cdot y_{1}(u-m+2) \cdots
  y_{m-1}(u-m+2) $ on both side. 
The other identities can be proved by a similar method. Here we list
${\cal M}$ and $(b,c)$ to be used in (\ref{jacobi}) and some other 
relations particularly needed.  
The equations (\ref{relation-sift}) and 
(\ref{relation-CR}) should be used as well. \\ 
(\ref{relation3}) for $m \in 2{\bf Z }_{\ge 0}+1 $: 
${\cal M}={\cal R}_{m+2}
          \left[
            \begin{array}{c}
             2\\
             2 
            \end{array}
      \right]
        (u-m)$ with $(b,c)=(1,m+1)$. \\       
(\ref{relation1}) for $m \in 2{\bf Z }_{\ge 0}+1 $: 
\begin{eqnarray}
{\cal M}=
  \left(
   \begin{array}{cccccc}
      0      & -1     & y_{1}   & y_{2}  & \ldots & y_{m}  \\ 
      1      &  0     & x_{1}   & x_{2}  & \ldots & x_{m}  \\
     -y_{1}      & -x_{1} & 0       & a_{12} & \ldots & a_{1m}  \\
         -y_{2}  & -x_{2} & -a_{12} & 0      & \ldots & a_{2m}  \\       
         \vdots  & \vdots & \vdots  & \vdots & \ddots & \vdots  \\
         -y_{m}  & -x_{m} & -a_{1m} & -a_{2m}& \ldots & 0
   \end{array}
  \right)
 (u-m+1) 
\end{eqnarray}
with $(b,c)=(1,2)$.\\
(\ref{relation4}): 
${\cal M}={\cal C}_{m+2}(u-m)$ with $(b,c)=(1,2)$.\\ 
(\ref{relation6}): 
${\cal M}={\cal C}_{m+2} 
    \left[
            \begin{array}{c}
             2\\
             2 
            \end{array}
      \right] (u-m)$ with $(b,c)=(1,m+1)$.\\ 
(\ref{relation7}): 
${\cal M}={\cal C}_{m+1}(u-m+1)$ with $(b,c)=(1,m+1)$.\\ 
(\ref{relation1}) for $m \in 2{\bf Z }_{\ge 0} $: 
${\cal M}={\cal S}_{m+1}(u-m+1)$ with $(b,c)=(1,m+1)$,  the relation 
(\ref{T_m^(r)}), (\ref{relation7}) and 
(\ref{relation3}).\\
(\ref{relation8}): 
${\cal M}={\cal C}_{m+2} 
    \left[
            \begin{array}{c}
             2\\
             2 
            \end{array}
      \right] (u-m-1)$ with $(b,c)=(1,2)$. 
\fullsqr  \\
We have presented similar relations for (\ref{T_m^(r-1)}) and 
(\ref{T_m^(r)}) in Appendix A . \\    
Proof of the Theorem \ref{main-th}.  
Eq. (\ref{t-sys1}) follows from Lemma\ref{formallemma} and (\ref{t-sys2}) 
from Lemma \ref{lemma-t-sys2.2} and (\ref{relation1}).
Eq. (\ref{t-sys3}) for $a=r$ is derived as follows.
Let 
${\cal M}= {\cal S}_{m+2}
          \left[
            \begin{array}{c}
             1 \\
             2
            \end{array}
      \right]
        (u-m)$
, then from (\ref{relation4}), (\ref{relation1}),
  (\ref{relation6}) and 
 (\ref{T_m^(a)}), we have 
\begin{eqnarray}
\fl  \det {\cal M}=(-1)^{m+1}T_{m}^{(r-1)}(u+1)T_{m+1}^{(r)}(u) , 
\  
 \det {\cal M}\left[
   \begin{array}{c}
        1 \\
        1 
   \end{array}
  \right]
  =(-1)^{m}T_{m}^{(r-1)}(u+1)T_{m}^{(r)}(u+1),   
        \nonumber \\
 \det {\cal M}\left[
   \begin{array}{c}
        m+1 \\
        m+1 
   \end{array}
  \right]
  =(-1)^{m}T_{m-1}^{(r-1)}(u)T_{m}^{(r)}(u-1), \nonumber \\  
  \det {\cal M}\left[
   \begin{array}{c}
        1 \\
        m+1 
   \end{array}
  \right]
  =(-1)^{m}T_{m}^{(r-1)}(u+1)T_{m-1}^{(r-1)}(u), \label{for jacobi1} \\ 
\fl  \det {\cal M}\left[
   \begin{array}{c}
        m+1 \\
        1 
   \end{array}
  \right]
  =(-1)^{m}T_{m}^{(r-2)}(u), \  
  \det {\cal M}\left[
   \begin{array}{cc}
        1 & m+1 \\
        1 & m+1
   \end{array}
  \right]
  =(-1)^{m-1}T_{m-1}^{(r-1)}(u)T_{m-1}^{(r)}(u). \nonumber 
\end{eqnarray}
Applying (\ref{jacobi}) for $b=1$ and $c=m+1$ to (\ref{for jacobi1}),
 we get (\ref{t-sys3}) for $a=r$. 
Eq. (\ref{t-sys3}) for $a=r-1$ is derived quite similarly. 
Let 
${\cal M}= {\cal S}_{m+3}
          \left[
            \begin{array}{cc}
             1 & 2 \\
             2 & 3 
            \end{array}
      \right]
        (u-m-2)$
, then from (\ref{relation1}), (\ref{relation8}), (\ref{T_m^(a)})  
and (\ref{relation7}), we have 
\begin{eqnarray}
\fl  \det {\cal M}=(-1)^{m+1}T_{m+1}^{(r-1)}(u)T_{m}^{(r)}(u+1) , \quad 
  \det {\cal M}\left[
   \begin{array}{c}
        1 \\
        1 
   \end{array}
  \right]
  =(-1)^{m}T_{m}^{(r-1)}(u+1)T_{m}^{(r)}(u+1),   
        \nonumber \\
     \det {\cal M}\left[
   \begin{array}{c}
        m+1 \\
        m+1 
   \end{array}
  \right]
  =(-1)^{m}T_{m}^{(r-1)}(u-1)T_{m-1}^{(r)}(u), \nonumber \\ 
  \det {\cal M}\left[
   \begin{array}{c}
        1 \\
        m+1 
   \end{array}
  \right]
  =(-1)^{m}T_{m-1}^{(r)}(u)T_{m}^{(r)}(u+1), \label{for jacobi4} \\ 
 \det {\cal M}\left[
   \begin{array}{c}
        m+1 \\
        1 
   \end{array}
  \right]
  =(-1)^{m}T_{m}^{(r-2)}(u),  \nonumber \\ 
  \det {\cal M}\left[
   \begin{array}{cc}
        1 & m+1 \\
        1 & m+1
   \end{array}
  \right]
  =(-1)^{m-1}T_{m-1}^{(r-1)}(u)T_{m-1}^{(r)}(u).
        \nonumber 
\end{eqnarray}
Applying (\ref{jacobi}) for $b=1$ and $c=m+1$ to (\ref{for jacobi4}),
 we have (\ref{t-sys3}) for $a=r-1$. \fullsqr \\  
Remark : Reflecting the Dynkin diagram symmetry of $D_{r}$ , 
similar relations to  Lemma\ref{main-lemma}, Theorem\ref{main-th} 
and Appendix A can be obtained by exchanging $T_{1}^{(r-1)}(u)$ 
 and $T_{1}^{(r)}(u)$.
\section{Discussion}
In this paper, we have given a new representation of the
solution to the $D_r$ $T$-system (1.1).
The key is the introduction of the auxiliary
dress function ${\cal T}^a$ (\ref{generating}) and the 
new functional relation (\ref{key}).
These are motivated from the analytic Bethe ansatz and 
lead to a different expression of the solution from the 
earlier one [KNH].

A similar analysis has been done in [KOS] for $B_r$ case.
There a more general class of transfer matrix spectra has been  
represented not only by determinants but also 
as summations over certain tableaux.
They are $B_r$ Yangian analogues of the semi-standard Young tableaux
for $sl(r+1)$.
There remains a problem to extend such an analysis to $D_r$ case.
So far we have only found a conjecture on the 
tableau sum representations of $T^{(r)}_m(u)$ and $T^{(r-1)}_m(u)$,
as stated below.

Consider an injection $\iota : Spin^\epsilon  \to J^r$,
sending $(\zeta_{1},\dots ,\zeta_{r})$
to $(i_1, \dots, i_k, \overline{j_{r-k}}, \ldots, \overline{j_1})$ 
such that $\zeta_{i_1}= \cdots = 
\zeta_{i_k}=+ $, $\zeta_{j_1}=\cdots = 
\zeta_{j_{r-k}}=- $, $1 \le  i_{1} <  \cdots < i_{k} \le r$ and
$1 \le  j_{1} <  \cdots < j_{r-k} \le r$ .
We shall write the components as
$\iota (\zeta)=(\iota (\zeta)_{1},
\ldots, \iota (\zeta)_{r})$.
For $\epsilon = \pm$ and $m \in {\bf Z}_{\ge 1}$ put
\begin{eqnarray}
\fl Spin_{m}^\epsilon=\{ (\zeta^{(1)}, \ldots, \zeta^{(m)}) 
\in (Spin^\epsilon)^m :
  \iota (\zeta^{(i)})_a \preceq \iota (\zeta^{(i+1)})_a ,
      \nonumber \\ 
      {\rm for }\,  1\le i \le m-1, \,  1 \le a \le r \} 
        \label{condition-spin}
\end{eqnarray}
This is well defined because the situations
$(\iota (\zeta^{(i)})_a, \iota (\zeta^{(i+1)})_a) = (r,\overline{r})$
and $(\overline{r}, r)$ never happen due to the parity constraint 
in (\ref{spin}).
In particular $Spin^\epsilon_1 = Spin^\epsilon$.
Now our conjecture reads
\begin{equation}
T_{m}^{(r+{\epsilon -1 \over 2})}(u)
=\sum_{(\zeta^{(1)}, \ldots, \zeta^{(m)}) \in Spin_{m}^\epsilon} 
                  \prod_{i=1}^{m} sp(\zeta^{(i)};u-m+2i-1).\label{conj}
\end{equation}
We have verified this for $4 \le r \le 6, \, 1 \le m \le 2$.

\ack 
The authors thank Toshiki Nakashima and Junji Suzuki 
for a helpful discussion.  
\appendix
\section{Other relations}
The following relations are valid.
\begin{equation}
 \fl T_{m}^{(r-1)}(u+1) T_{m}^{(r)}(u-1)=
    \left\{
          \begin{array}{ll}
            \det[{\cal C}_{m+2}
            \left[
              \begin{array}{cc}
               1 & 2 \\
               1 & m+2
              \end{array}
            \right]
            (u-m)] & {\rm for} \quad  m \in 2{\bf Z }_{\ge 0} \\ 
            \det[{\cal S}_{m+2} 
            \left[
              \begin{array}{c}
                2 \\
                m+2
              \end{array}
            \right]
            (u-m)] 
            & {\rm for} \quad  m \in 2{\bf Z }_{\ge 0}+1,
          \end{array}
        \right.
   \label{relation2}
\end{equation}
\begin{eqnarray}
 \fl T_{m-1}^{(r-1)}(u) T_{m}^{(r)}(u+1)=
    \left\{
          \begin{array}{ll}
            \det[{\cal S}_{m+2}
            \left[
              \begin{array}{cc}
                2 & m+2 \\
                1 & 2
              \end{array}
            \right]
            (u-m)] 
            & {\rm for} \quad  m \in 2{\bf Z }_{\ge 0} \\
             \det[{\cal S}_{m+1}
            \left[
              \begin{array}{c}
               m+1 \\
               1
              \end{array}
            \right]
            (u-m+2)] & {\rm for} \quad  m \in 2{\bf Z }_{\ge 0} +1,
      \end{array}
    \right.
 \label{relation5}
\end{eqnarray}

\begin{eqnarray}
\fl  T_{m+1}^{(r-\delta_{m})}(u) T_{m}^{(r-\delta_{m})}(u+1)=
    \det[{\cal S}_{m+2}
           \left[
            \begin{array}{c}
             2 \\
             1
            \end{array}
           \right]
           (u-m)],
 \label{relation9}
\end{eqnarray}
\begin{eqnarray}
\fl T_{m+1}^{(r-\delta_{m-1})}(u) T_{m}^{(r-\delta_{m-1})}(u+1)=
    \det[{\cal S}_{m+3} 
           \left[
            \begin{array}{cc}
             2 & 3 \\
             1 & 2
            \end{array}
           \right]
           (u-m-2)],
 \label{relation10}
\end{eqnarray}

\begin{eqnarray}
 \fl T_{m}^{(r-1)}(u-1) T_{m}^{(r)}(u+1)=
    \left\{
          \begin{array}{l}
            \det[{\cal C}_{m+3}
            \left[
              \begin{array}{ccc}
                1 & 2 & m+3 \\
                1 & 2 & 3
              \end{array}
            \right]
            (u-m-2)] \\ 
            \qquad {\rm for} \quad  m \in 2{\bf Z }_{\ge 0} \\
             \det[{\cal S}_{m+3} 
            \left[
              \begin{array}{cc}
               2 & m+3 \\
               2 & 3
              \end{array}
            \right]  
            (u-m-2)] \\ \qquad {\rm for} \quad  m \in 2{\bf Z }_{\ge 0} +1,
      \end{array}
    \right.
 \label{relation11}
\end{eqnarray}

\begin{eqnarray}
 \fl T_{m+1}^{(r-1)}(u) T_{m-1}^{(r)}(u)=
    \left\{
          \begin{array}{ll}
             \det[{\cal S}_{m+3}
            \left[
              \begin{array}{cc}
               2 & m+3 \\
               2 & 3
              \end{array}
            \right]
            (u-m-2)] 
            & {\rm for} \quad  m \in 2{\bf Z }_{\ge 0} \\
              \det[{\cal C}_{m+3} 
            \left[
              \begin{array}{ccc}
                1 & 2 & m+3 \\
                1 & 2 & 3
              \end{array}
            \right]
            (u-m-2)] \\ 
            \qquad {\rm for} \quad  m \in 2{\bf Z }_{\ge 0} +1,
      \end{array}
    \right.
 \label{relation12}
\end{eqnarray}

\begin{eqnarray}
 \fl T_{m-1}^{(r-1)}(u) T_{m+1}^{(r)}(u)=
    \left\{
          \begin{array}{ll}
            \det[{\cal S}_{m+2} 
            \left[
              \begin{array}{c}
               m+2 \\
               2
              \end{array}
            \right]
            (u-m)] 
            & {\rm for} \quad  m \in 2{\bf Z }_{\ge 0} \\
              \det[{\cal C}_{m+2} 
            \left[
              \begin{array}{cc}
                1 & m+2 \\
                1 & 2
              \end{array}
            \right]
            (u-m)] 
            & {\rm for} \quad  m \in 2{\bf Z }_{\ge 0} +1.
      \end{array}
    \right.
 \label{relation13}
\end{eqnarray}
Proof. 
The proof is done in the same way as Lemma\ref{main-lemma}. 
Here we list ${\cal M}$ and $(b,c)$ to be used in (\ref{jacobi}) and some other 
relations particularly needed.  
The equations (\ref{relation-CR}) and (\ref{relation-sift}) 
should also be used when necessary. \\
(\ref{relation2}) ; for $m \in 2{\bf Z }_{\ge 0} $ : 
${\cal M}={\cal C}_{m+2} 
          \left[
            \begin{array}{c}
             1 \\
             1  
            \end{array}
      \right]
        (u-m)$ with $(b,c)=(1,m+1)$, \\ 
 for $m \in 2{\bf Z }_{\ge 0}+1 $ : 
${\cal M}={\cal S}_{m+2}(u-m)$ with $(b,c)=(2,m+2)$ .\\ 
(\ref{relation5}) ; for $m\in 2{\bf Z }_{\ge 0}$ : 
${\cal M}={\cal R}_{m+2} 
    \left[
            \begin{array}{c}
             2\\
             2 
            \end{array}
      \right] (u-m)$ with $(b,c)=(1,m+1)$ , \\ 
for $m\in 2{\bf Z }_{\ge 0}+1$ : 
${\cal M}={\cal R}_{m+1}(u-m+2)$ with $(b,c)=(1,m+1)$ . \\ 
(\ref{relation9}) : 
${\cal M}={\cal S}_{m+2}(u-m)$ with $(b,c)=(1,2)$,
   the relations (\ref{relation1}), (\ref{relation4}). \\
(\ref{relation10}) : 
${\cal M}={\cal S}_{m+3} 
    \left[
            \begin{array}{c}
             2\\
             2 
            \end{array}
      \right] (u-m-2)$ with $(b,c)=(1,2)$, 
      the relations (\ref{relation1}), (\ref{relation8}).\\
 (\ref{relation11}) ; for $m \in 2{\bf Z }_{\ge 0} $ : 
${\cal M}={\cal C}_{m+3} 
          \left[
            \begin{array}{cc}
             1 & 2 \\
             1 & 2  
            \end{array}
      \right]
        (u-m-2)$ with $(b,c)=(1,m+1)$,\\ 
for $m \in 2{\bf Z }_{\ge 0}+1 $  : 
${\cal M}={\cal C}_{m+3} 
          \left[
            \begin{array}{c}
             2 \\
             2 
            \end{array}
      \right]
        (u-m-2)$ with $(b,c)=(2,m+2)$.  \\ 
(\ref{relation12}) ; for $m \in 2{\bf Z }_{\ge 0} $ : 
${\cal M}={\cal C}_{m+3} 
          \left[
            \begin{array}{c}
             2 \\
             2  
            \end{array}
      \right]
        (u-m-2)$ with $(b,c)=(2,m+2)$,\\ 
for $m \in 2{\bf Z }_{\ge 0}+1 $ : 
${\cal M}={\cal C}_{m+3} 
          \left[
            \begin{array}{cc}
             1 & 2\\
             1 & 2 
            \end{array}
      \right]
        (u-m-2)$ with $(b,c)=(1,m+1)$.\\ 
(\ref{relation13}) ; for $m \in 2{\bf Z }_{\ge 0} $ : 
${\cal M}={\cal S}_{m+2}(u-m)$ with $(b,c)=(2,m+2)$,\\ 
for $m \in 2{\bf Z }_{\ge 0}+1 $ : 
${\cal M}={\cal C}_{m+2}
          \left[
            \begin{array}{c}
             1 \\
             1  
            \end{array}
      \right]
        (u-m)$ with $(b,c)=(1,m+1)$. 
\fullsqr  \\ 
\section*{References}
\begin{harvard}

       \item[] [AL]
       Ablowitz M J and Ladik F J 1976 {\it Stud. Appl. Math.} {\bf 55}
       213; 1977 {\it Stud. Appl. Math.} {\bf 57} 1

       \item[] [DJM]
       Date E, Jimbo M and Miwa T 1982 \JPSJ {\bf 51} 4116;4125;
       1983 \JPSJ {\bf 52} 388;761;766

        \item[] [H]
        Hirota R 1977 \JPSJ {\bf 43} 1424; 1978 \JPSJ {\bf 45} 321;
        1981 \JPSJ {\bf 50} 3785; 1987 \JPSJ {\bf 56} 4285
    
        \item[] [KLWZ] Krichever I, Lipan O, Wiegmann P and Zabrodin A  
        {\it Quantum Integrable Models and Discrete Classical Hirota
             Equations}, preprint ESI 330 (1996)        
    
        \item[] [KNH]
        Kuniba A, Nakamura S and Hirota R  1996 \JPA{\bf 29} 1759
        
        \item[] [KNS]
        Kuniba A, Nakanishi T and Suzuki J 1994 {\it Int. J. Mod. Phys.}
         {\bf A9} 5215 

        \item[] [KOS]
        Kuniba A, Ohta Y and Suzuki J  1995 \JPA{\bf 28} 6211
        
        \item[] [KS]
        Kuniba A and Suzuki J 1995 {\it Commun. Math. Phys.} {\bf 173} 225 

        \item[] [LS]
        Leznov A N and Saveliev M V 1979 {\it Lett. Math. Phys.} {\bf 3} 489

        \item[] [MOP]
        Mikhailov A V, Olshanetsky M A and Perelomov A M 1981 
        {\it Commun. Math. Phys.} {\bf 79} 473
        
        \item[] [R]
        Reshetikhin N Yu 1983 {\it Sov. Phys.-JETP} {\bf 57} 691 
                
\end{harvard}
\end{document}